\documentclass[12pt,preprint]{aastex}
\shorttitle{Population of Abell Clusters}
\shortauthors{Barkhouse et al.}
\begin{document}
\title{The Galaxy Population of Low-Redshift Abell Clusters}
\author{Wayne A. Barkhouse,\altaffilmark{1,4} H.K.C. Yee,
\altaffilmark{2,4} and Omar L\'{o}pez-Cruz\altaffilmark{3,4}}
\altaffiltext{1}{Department of Physics and Astrophysics, University of 
North Dakota, Grand Forks, 
ND 58202; email: wayne.barkhouse@und.nodak.edu}
\altaffiltext{2}{Department of Astronomy and Astrophysics, University 
of Toronto, Toronto, ON, Canada, M5S 3H8; email: hyee@astro.utoronto.ca}
\altaffiltext{3}{Instituto Nacional de Astrof\'{i}sica, Optica y 
Electr\'{o}nica, Tonantzintla, Pue., M\'{e}xico; email: omarlx@inaoep.mx}
\altaffiltext{4}{Visiting Astronomer, Kitt Peak National Observatory. 
KPNO is operated by AURA, Inc.\ under contract to the National Science
Foundation.}

\begin{abstract}

We present a study of the luminosity and color properties of galaxies 
selected from a sample of 57 low-redshift Abell clusters. We utilize the 
non-parametric dwarf-to-giant ratio (DGR) and the blue galaxy fraction 
($f_{b}$) to investigate the clustercentric radial-dependent changes 
in the cluster galaxy population. Composite cluster samples are combined 
by scaling the counting radius by $r_{200}$ to minimize radius selection 
bias. The separation of galaxies into a red and blue population was achieved 
by selecting galaxies relative to the cluster color-magnitude relation. The 
DGR of the red and blue galaxies is found to be independent of cluster 
richness ($B_{gc}$), although the DGR is larger for the blue population at 
all measured radii. A decrease in the DGR for the red and red+blue galaxies is 
detected in the cluster core region, while the blue galaxy DGR is nearly 
independent of radius. The $f_{b}$ is found not to correlate with $B_{gc}$; 
however, a steady decline toward the inner-cluster region is observed for 
the giant galaxies. The dwarf galaxy $f_{b}$ is approximately constant 
with clustercentric radius except for the inner cluster core region where 
$f_{b}$ decreases. The clustercentric radial dependence of the DGR and the 
galaxy blue fraction, indicates that it is unlikely that a simple scenario 
based on either pure disruption or pure fading/reddening can describe the 
evolution of infalling dwarf galaxies; both outcomes are produced by the 
cluster environment.

\end{abstract}

\section{Introduction}

A fundamental goal in the study of galaxy clusters is to understand the role of 
environment on galaxy formation and evolution. 
The well-established morphology--density relation 
\citep[e.g.,][]{Dressler80,Dressler97,Thomas06} highlights the impact of 
location on the properties of cluster galaxies: the cores of rich clusters 
are inundated with early-type galaxies while the outskirts contain a large 
fraction of late-type systems \citep[for example, see][]{Abraham96,Morris98,
Treu03,Smith06}. 

A theoretical understanding of the morphology--density relation has 
focused on the influence of dynamical factors on the formation of early-type 
galaxies in high-density regions via the merger of late-type galaxies 
\citep[e.g.,][]{Okamoto01}. Mergers are expected to occur with greater 
ease in areas where the galaxy velocity dispersion is low \citep[e.g.,][]{Merritt84}. 
This is in contrast to the high-velocity dispersion regions found at the center of 
rich clusters \citep[e.g.,][]{Rood72,Kent82,Dubinski98}. To solve this 
apparent contradiction with the observed morphology--density relation, it is 
hypothesized that the transformation of late- into early-type galaxies 
occurred in group-like environments with a low-velocity dispersion, as 
clusters assembled from the gravitational infall of matter 
\citep[e.g.,][]{Roos79,McIntosh04}. Once a cluster has formed, other 
dynamical processes, such as galaxy harassment and ram pressure 
stripping, will determine the current morphological makeup of the cluster 
galaxy population \citep{Moore96,Abadi99,Quilis00,Boselli06}.

In general, low-redshift clusters contain two major galaxy populations; a 
red, evolved, early-type component that dominates the central cluster region, 
and a blue, late-type population which has undergone relatively recent star 
formation and is most prominent in the outskirts of clusters 
\citep[for example, see][]{Abraham96,Wake05,Wolf07}. This has led to the 
hypothesis that clusters are built-up gradually from the infall of 
field galaxies \citep[e.g.,][]{Ellingson01,Treu03,McIntosh04}.

Understanding how different sub-populations of galaxies evolve in clusters 
can be further elucidated by subdividing the cluster galaxy population with 
respect to luminosity and color. \citet[][hereafter B07]{Barkhouse07} 
demonstrated that, in general, the faint-end slope of the cluster luminosity 
function (LF) becomes steeper with increasing clustercentric distance. In 
this paper we examine the radial dependence of galaxy luminosity by 
measuring the dwarf-to-giant ratio (DGR) as a function of clustercentric 
radius. Unlike the galaxy LF, the non-parametric DGR provides a robust 
measure of the relative fraction of faint-to-bright galaxies without 
assuming a specific functional form for the LF. 
In \citet{Lopez04} we presented evidence for a blueward shift in the 
color-magnitude relation (CMR) with increasing clustercentric radius, 
utilizing the same cluster sample as this paper. To quantify changes in 
galaxy color as a function of clustercentric radius, we search for 
radius-dependent changes in the blue galaxy fraction ($f_{b}$).

This paper is the third in a series resulting from a large multi-color 
imaging survey of low-redshift Abell galaxy clusters. This paper is 
organized as follows. In \S 2 we briefly summarize the sample selection 
criteria, observations, and photometric reductions. In \S 3 we examine 
the dwarf-to-giant ratio, while the blue galaxy fraction is explored in \S 4. 
We compare our findings with published results in \S 5 and discuss our 
findings in \S 6. Finally we summarize our conclusions in \S 7.

Further details regarding sample selection, observations, image 
preprocessing, catalogs, and finding charts can be found in 
\citet{Lopez97}, \citet{Barkhouse03}, \citet{Lopez04}, and O. L\'{o}pez-Cruz 
et al. (2009, in preparation). The investigation of the color-magnitude relation 
for this cluster sample is presented in Paper I \citep{Lopez04}, while the 
galaxy cluster LFs are presented in Paper II (B07).

Due to the low redshift nature of our cluster sample ($z< 0.2$), the effects 
of curvature and dark energy are negligible. To allow a direct comparison 
with previous studies, we use 
$H_{0}=50\,h_{50}~\mbox{km}~\mbox{s}^{-1}~\mbox{Mpc}^{-1}$ and $q_{0}=0$, 
unless otherwise indicated.

\section{Observations and Data Reductions}

We present in this section a brief summary of the cluster sample selection 
criteria, observations, data reductions, and photometric measurements. We 
refer the reader to \citet{Lopez97}, \citet{Barkhouse03}, \citet{Lopez04}, 
and B07 for further details.  

The Abell clusters in our sample are selected mainly from the catalog of 
{\it Einstein}-detected bright X-ray clusters compiled by \citet{Jones99}. 
This sample includes 47 clusters observed at KPNO using the 0.9 m telescope 
plus the T2kA ($2048 \times 2048$ pixels; $0.68\arcsec~\mbox{pixel}^{-1}$) CCD 
detector \citep[the LOCOS sample;][O. L\'opez-Cruz et al. 2009, in 
preparation]{Lopez97,Yee99,Lopez01}. These clusters 
were chosen to be at high galactic latitude ($|b| \geq 30\degr$) and within 
the redshift range of $0.04\leq z\leq 0.20$. To probe the low redshift regime 
($0.02\leq z\leq 0.04$), eight clusters imaged using the KPNO 0.9 m + MOSAIC 
8K camera ($8192\times 8192$ pixels; $0.423\arcsec~\mbox{pixel}^{-1}$) from 
\citet{Barkhouse03} were incorporated in our sample. In addition, two clusters 
from \citet{Brown97} using the same instrumental setup as the LOCOS sample 
and selection criteria as the mosaic data are included.

The integration times for our 57-cluster sample varies from 250 to 9900 s, 
depending on the filter ($B$ or $R_{C}$) and the redshift of the cluster. 
For this study we use 
a total of five control fields, which were chosen at random positions on the 
sky at least $5\degr$ away from the clusters in the sample. These control 
fields were observed using the MOSAIC camera to a comparable depth and 
reduced in the same manner as the cluster data.

Preprocessing of the images was done using IRAF and photometric 
reductions were carried out using the program PPP \citep[Picture Processing 
Package;][]{Yee91}, which includes algorithms for performing automatic object 
finding, star/galaxy classification, and total magnitude determination. 
Galaxy colors are measured using fixed apertures on the images of each 
filter, sampling identical regions of galaxies in different 
filters. Instrumental magnitudes are calibrated to the Kron-Cousins system by 
observing standard stars from \citet{La92}.

The 100\% completeness limit of each field was set at 1.0 mag brighter than 
the magnitude of a stellar object with a brightness equivalent to having 
a S/N=5 in an aperture of $2\arcsec$ \citep[see][]{Yee91}. Extinction values 
for each cluster are taken from the maps and tables of \citet{Burstein82} and 
\citet{Burstein84}.


\section{The Dwarf-to-Giant Ratio}

In B07 we presented the composite cluster 
LF for the sample of 57 low-redshift clusters utilized in this paper. 
We found that, in general, the slope of the faint-end of the LF rises with 
increasing clustercentric radius. To further expand upon these findings, 
we have divided the cluster galaxy population into two sub-samples based on 
luminosity. Following the nomenclature used by previous studies 
\citep[e.g.,][]{Lopez97,Driver98,Popesso06}, we classify galaxies brighter 
than $M_{R_{c}}= -20$ as ``Giants'' and those having 
$-19.5\le M_{R_{c}} \le -17.0$ as ``Dwarfs''. We note that $M^{\ast}_{R_{c}}$ 
calculated from these data (see B07) is $M^{\ast}_{R_{c}}\sim -22.3$. The 
measurement of $R_{c}$, including applied k-corrections, is described in 
B07. These magnitude cuts allow us to maximize the number of galaxies to 
improve statistical inferences while ensuring that incompleteness effects at 
the faint-end are negligible. We construct the cluster DGR by dividing the 
number of background-corrected dwarfs by the number of background-corrected 
giants. Only clusters 100\% photometrically complete to $M_{R_{c}}=-17$ are 
included in the construction of the DGR. Using this definition, we can employ 
the DGR to explore the change in the relative fraction of dwarfs and giants 
with respect to various cluster characteristics.

The primary advantage of using the DGR is that it provides a 
non-parametric measurement of the relative change in the number of giant 
and dwarf galaxies that is independent of the functional form selected for 
the cluster LF \citep[e.g., a Schechter function;][]{Schechter76}. For 
example, a steepening of the faint-end slope of the LF with increasing 
clustercentric radius would be characterized by an increase in the DGR
\citep{Driver98}. Given the degree of degeneracy between the parameters in 
a Schechter function fit (e.g., $M^{\ast}$ and $\alpha$; B07), the DGR 
yields additional insights on the luminosity distribution of cluster galaxies 
and thus complements the LF.

\subsection{DGR versus Richness}

\begin{figure}
\figurenum{1}
\epsscale{0.8}
\plotone{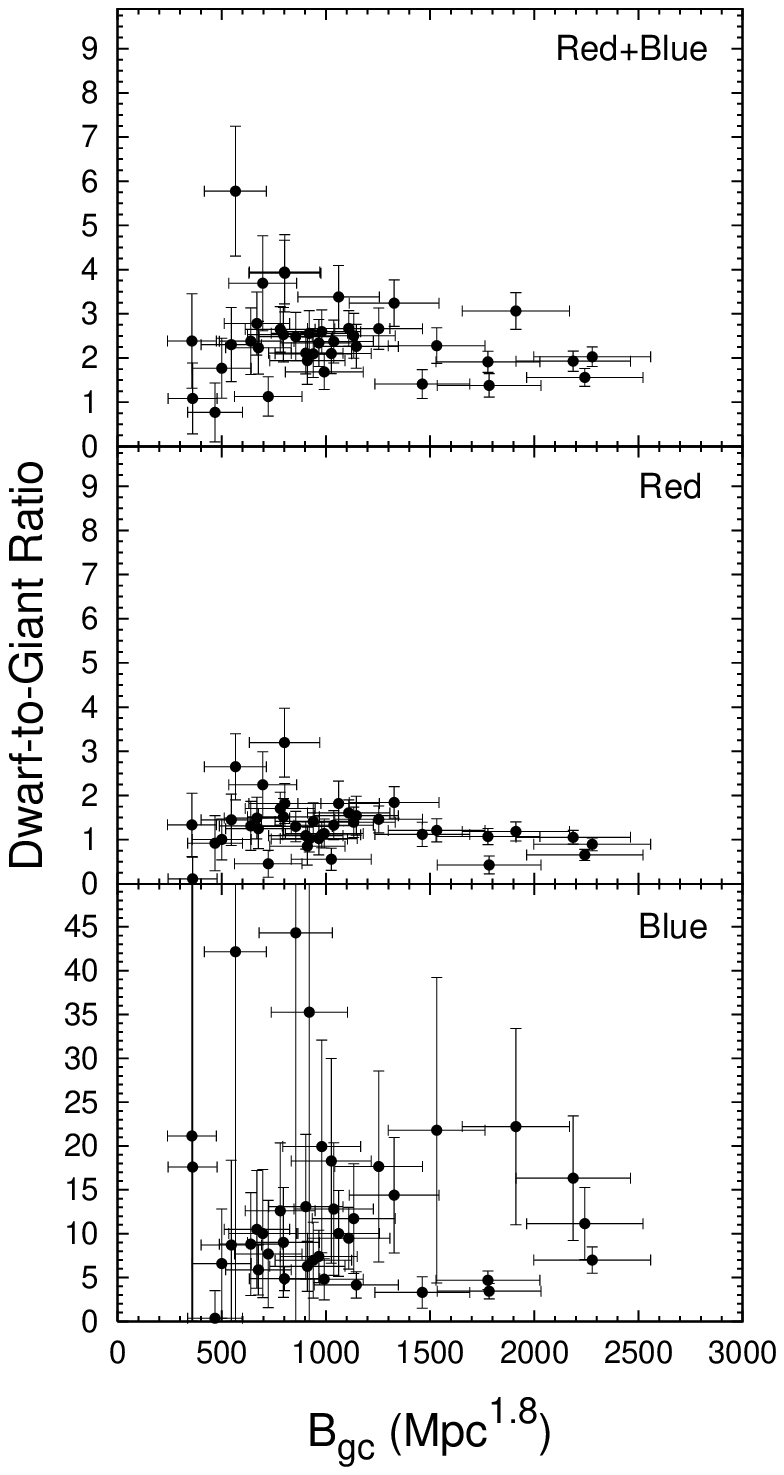}
\caption{
Comparison of the DGR with cluster richness for the red+blue ({\it top}), 
red ({\it middle}), and blue ({\it bottom}) galaxy populations that are 
photometrically complete to $M_{R_{c}}=-17$. The DGR is calculated for 
galaxies measured within $(r/r_{200})\leq0.4$.} 
\label{All-DGR-Bgc}
\end{figure}

To examine for a possible correlation of the DGR with cluster richness, we 
plot in Figure~\ref{All-DGR-Bgc} the DGR versus the richness parameter 
$B_{gc}$ for 39 clusters that are 100\% photometrically complete to 
$M_{R_{c}}=-17.0$ (DGR uncertainties are calculated assuming Poisson statistics 
and quadrature summation). The cluster optical richness is parameterized 
by $B_{gc}$, which is a measure of the cluster center--galaxy correlation 
amplitude (Yee \& L\'opez-Cruz 1999; Yee \& Ellingson 2003; B07). In 
Figure~\ref{All-DGR-Bgc} galaxies are selected within $(r/r_{200})= 0.4$ 
($\sim 900~h^{-1}_{50}~\mbox{kpc}$ for the sample average) of each cluster 
center for the red+blue, red, and blue cluster galaxy populations. The 
red+blue cluster galaxy sample is compiled by including galaxies that have 
been culled of systems redder than the CMR for each individual cluster (for 
further details, see B07). The center of each cluster is chosen as the position of the 
brightest cluster galaxy (BCG) or, when some doubt exists, the nearest 
bright early-type galaxy to the X-ray centroid. The value of $r_{200}$, 
an approximate measure of the virial radius \citep{Cole96}, is derived 
from the relationship between $r_{200}$ and $B_{gc}$ as described in 
B07 (see their equation 7), and is used as a scaling factor to minimize 
radial sampling bias due to variations in cluster richness
\citep{Christlein03,Hansen05,Popesso06}. The $r_{200}$ measurements  
for our cluster sample are tabulated in Table 1 of B07 ($r_{200}$ uncertainties are based 
on the 15\% rms scatter in the derived value of $r_{200}$ as explained in B07). 
The radial sampling criterion of $(r/r_{200})=0.4$ was chosen to maximize 
the number of clusters in our sample that are photometrically complete to 
$M_{R_{c}}=-17$, and thus provide a large number of net galaxy counts to 
minimize the statistical uncertainty in the DGR.

Analysis of Figure~\ref{All-DGR-Bgc} for the red+blue galaxies (top panel) indicates 
that there is no significant correlation between the DGR and cluster richness. 
A Kendell's $\tau$ statistic \citep{Press92} yields a 49\% probability that the 
DGR and $B_{gc}$ are correlated. When not employing a dynamical counting radius 
(e.g. $r_{200}$), the mean and dispersion of the DGR vs. $B_{gc}$ is increased 
significantly, indicating that the DGR is dependent on clustercentric radius 
(see \S 3.2 for further discussions).  

We also plot in Figure~\ref{All-DGR-Bgc} the DGR versus cluster richness for 
the red (middle panel) and blue (bottom panel) galaxy populations. The 
color-selection of galaxies into red and blue samples is described in 
detail in B07. To briefly summarize; the red galaxies are chosen 
if they are located within $\pm 0.22$ mag ($3\sigma$) of the cluster CMR, 
while blue galaxies are selected from the area of the color-magnitude 
diagram that is blueward of the region used to select the red galaxies 
(see Figures 5 and 8 from B07). As noted for the combined red+blue galaxy 
population, the red and blue galaxies have been background-corrected 
using the exact same color-selection criteria as for the cluster galaxies. 

Similar to the red+blue sample, we find no significant correlation between 
the DGR and richness when considering the red and blue galaxies separately. 
A Kendell's $\tau$ statistic indicates a 67\% probability that the red DGR 
is correlated with $B_{gc}$; while for the blue DGR, there is only a 
3\% probability that these quantities are correlated. 

For the red+blue population we find a mean DGR of $2.41^{+1.28}_{-0.47}$, 
where the uncertainties bracket the interval containing 68\% of the data points 
about the mean. For the red and blue samples we have 
$1.30^{+0.41}_{-0.38}$ and $11.20^{+11.02}_{-4.61}$, respectively. The 
dominance of the dwarfs for the blue galaxy population compared to those in 
the red galaxy sample, is consistent with the findings 
from B07 in the sense that the blue LFs were found to have a steeper faint-end 
slope than the red sequence LFs. It is interesting to note that the blue DGR 
has a larger range of values and dispersion than the red DGR. This is an 
indication that blue galaxies in clusters have a larger variance in their 
properties and states of evolution compared to the red galaxies, which can be 
seen as primarily dominated by the end products of galaxy evolution and infall 
process.

\subsection{DGR versus Clustercentric Radius}

\begin{figure}
\figurenum{2}
\epsscale{1.0}
\plotone{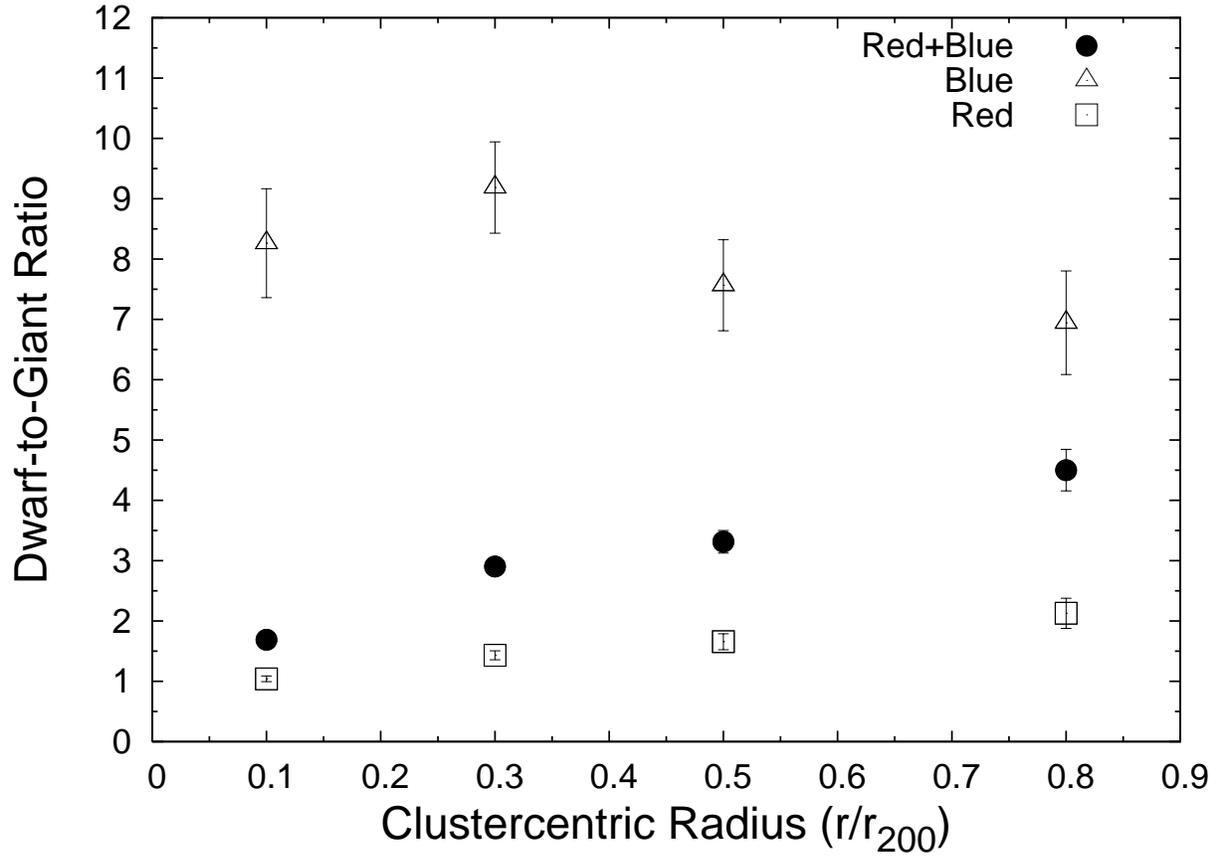}
\caption{DGR as a function of clustercentric radius for the red+blue 
(filled circles), red (open squares), and blue (open triangles) 
cluster galaxy populations.}
\label{RedBlue-Radius}
\end{figure}

In B07 we presented evidence for a steepening of the LF faint-end slope 
towards the cluster outskirts. To investigate this correlation 
further, we plot in Figure~\ref{RedBlue-Radius} the DGR for the red+blue, red, 
and blue galaxy populations as a function of clustercentric radius. 
The DGR is plotted at the mid-point for the following annuli; 
$(r/r_{200})\leq 0.2$, $0.2\leq (r/r_{200})\leq 0.4$, 
$0.4\leq (r/r_{200})\leq 0.6$, and $0.6\leq (r/r_{200})\leq 1.0$. The DGR 
is constructed by stacking cluster galaxies appropriate for each radial 
bin, and the uncertainty is derived from the standard deviation assuming 
Poisson statistics. Due to the variation in spatial imaging coverage, the 
number of clusters contributing to each radial bin is not equal. For all 
radius-dependent analysis in this study, we only include clusters that have 
complete spatial coverage for the indicated clustercentric radius.

Examination of Figure~\ref{RedBlue-Radius} shows that the DGR increases with 
radius for both the red and red+blue galaxy populations. For the red+blue 
galaxy sample the DGR is 2.7 times larger for the outer-most 
annulus as compared to the inner-most radial bin ($8.0\sigma$ difference). 
For the red galaxies the DGR is $2.0$ times larger in the outer-most annulus 
as compared to the central region (significant at the 
$4.3\sigma$ level). The DGR for the blue galaxies is approximately constant 
with perhaps even a possible decreasing trend with radius. We note that the 
uncertainties for the blue DGR are $\sim5$ times greater than those for the 
red and red+blue samples. This is a direct result of a small background-corrected 
blue giant count. A weighted linear least-squares fit to the blue galaxy data 
indicates that the radial slope of the DGR is different from zero at the 
$1.7\sigma$ level. Although the error bars are relatively large, the distribution of 
the blue DGR suggests a mild rising trend with {\it decreasing} radius. The rise in 
the DGR with increasing radius for the red+blue galaxy sample is a reflection 
of the increasing dominance of blue galaxies with radius. This result is in 
agreement with Figures 4 and 7 from B07 where the red and red+blue galaxy 
populations were found to have a rising faint-end slope with increasing 
clustercentric radius. The faint-end slope of the blue galaxy LF depicted in 
Figure 10 from B07 indicates a much weaker dependence on radius.

Figure~\ref{RedBlue-Radius} also shows that for all radii depicted, the 
DGR is larger for the blue than for the red galaxy 
sample. This finding is also supported by the comparison of red and blue LFs 
presented in Figure 11 of B07, and is a result of the larger contribution 
of luminous galaxies to the red sample as compared to the blue population. 


\section{The Blue Galaxy Fraction}

A comparison of the number of red and blue galaxies gives a rough indication 
of the relative mixture of early- and late-type systems. Due to the poor 
seeing of our imaging data (fwhm$\sim 1.5^{\prime\prime}$), morphological 
classification of galaxies with magnitudes near the completeness limit 
are not reliable. We therefore elect to use a broad-band color selection 
technique, such as the fraction of blue galaxies ($f_{b}$; the number of blue 
galaxies divided by the number of red+blue galaxies), to glean some information 
about the galaxy population makeup of our sample. We note that $f_{b}$ is 
constructed from background-corrected galaxy counts, and only includes clusters 
that are 100\% photometrically complete for the indicated magnitude range.''

\subsection{Richness Dependence of $f_{b}$}

\begin{figure}
\figurenum{3}
\epsscale{1.0}
\plotone{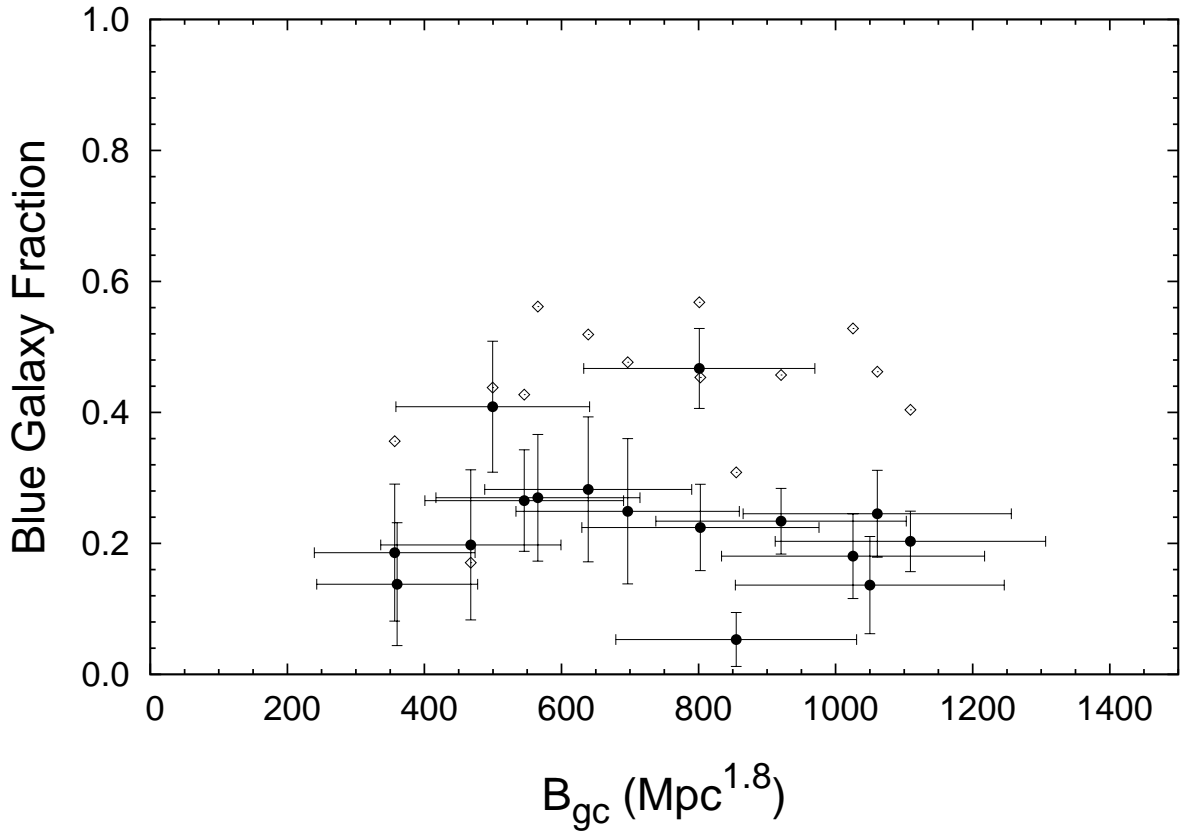}
\caption{$f_{b}$ as a function of cluster richness ($B_{gc}$) for the cluster 
galaxy population complete to $M_{R_{c}}=-19$ (filled circles) and 
$M_{R_{c}}=-17$ (open diamonds). Error bars for the open symbols are 
similar to the filled circles and have been omitted for clarity. Galaxies 
have been measured within a clustercentric radius of $(r/r_{200})=0.8$.}
\label{fb-Bgc}
\end{figure}

In Figure~\ref{fb-Bgc} we present $f_{b}$ versus cluster richness 
for galaxies selected within $(r/r_{200})=0.8$ of each cluster center. 
To search for luminosity-related dependencies in the 
distribution of $f_{b}$ with cluster richness, we plot $f_{b}$ for galaxies 
with $M_{R_{c}}\leq -17$ (14 clusters; open symbols) and $M_{R_{c}}\leq -19$ 
(16 clusters; filled symbols). We note that the magnitude limit utilized by 
\citet{Butcher84} to characterize the cluster blue fraction corresponds to 
$M_{R_{c}}\sim-20.5$ using our adopted cosmology and filter \citep{Fukugita95}. 
In order to minimize redshift bias (i.e., the dominance in our sample of rich clusters 
at high redshift), we selected a redshift range ($z\leq 0.094$) such that 
a fair representation of cluster richness is available. Using clusters with $z\leq 0.094$, 
we find for the bright sample ($M_{R_{c}}\leq -19$), $\overline{f_{b}}=0.23\pm 0.08$, 
while for the deep sample ($M_{R_{c}}\leq -17$) we measure $\overline{f_{b}}=0.44\pm 0.10$ 
(uncertainties are calculated assuming Poisson statistics). Examination of Figure~\ref{fb-Bgc} 
reveals that there is no significant correlation between $f_{b}$ and $B_{gc}$ 
for either the deep or bright sample. A Kendell's $\tau$ statistic yields a 59\%(47\%) 
probability of a correlation for the deep(bright) sample. Using $(r/r_{200})=1$ as our counting 
aperture also yields no significant correlation between $f_{b}$ and $B_{gc}$. Using the equivalent 
Butcher \& Oemler magnitude counting limit of $M_{R_{c}}\sim-20.5$, we find 
$\overline{f_{b}}=0.16\pm 0.08$. A Kendell's $\tau$ statistic gives a 75\% probability of a 
correlation between $f_{b}$ and $B_{gc}$. Thus no significant correlation between $f_{b}$ and cluster 
richness is discernible when counting galaxies within an equivalent dynamical radius. 
This indicates that the galaxy population in clusters is not dependent on cluster 
richness if a dynamics-dependent radius is used in sampling.

\subsection{Radial and Magnitude Dependence of $f_{b}$}

\begin{figure}
\figurenum{4}
\epsscale{1.0}
\plotone{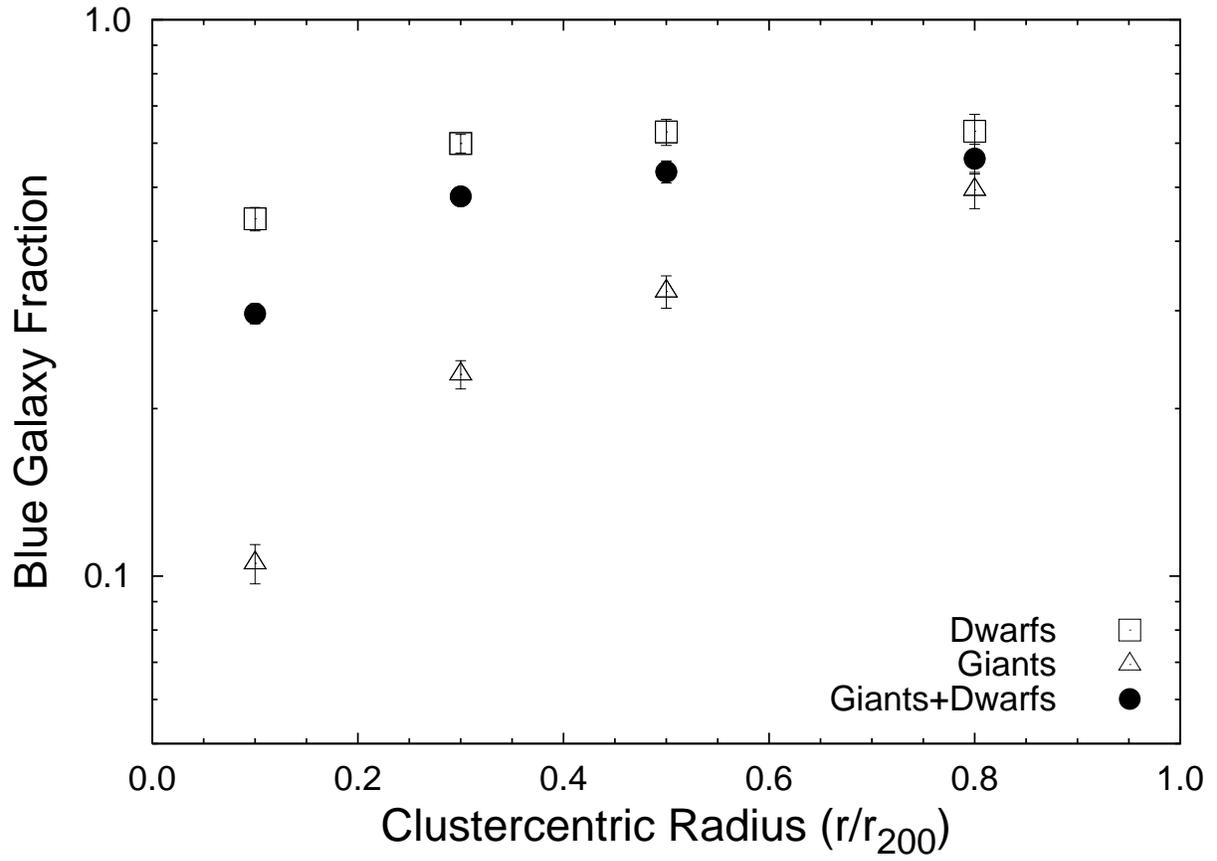}
\caption{$f_{b}$ as a function of clustercentric radius for the 
giants (open triangles), dwarfs (open squares), and giants+dwarfs (filled circles) 
cluster galaxy populations.}
\label{fb-Radius}
\end{figure}

To search for a possible radial dependence of $f_{b}$, we plot in 
Figure~\ref{fb-Radius} the $f_{b}$ in concentric annuli versus clustercentric 
radius for the dwarfs (open squares), giants (open triangles), and giants+dwarfs 
(filled circles) samples. Several aspects of the galaxy $f_{b}$ are apparent: 
a) at any radius the dwarf galaxies have a greater $f_{b}$ than the giants, b) 
the giant $f_{b}$ increases approximately monotonically with increasing radius, with 
a five-fold increase from the inner to the outer radial bin ($10\sigma$ 
difference), and c) the dwarf $f_{b}$, while decreasing by a factor of 
$\sim1.4$ at the inner-most radius, stays approximately constant at radii 
greater than $(r/r_{200})\sim0.2$.

To determine the relative change in the red and blue galaxies for 
the two inner-most radial bins, we restrict our cluster sample to include 
only those clusters that contribute to both annuli. Using this common 
cluster sample we find that there is a 53\% decrease 
in the net number of blue dwarfs when comparing the 
$0.2\leq (r/r_{200})\leq 0.4$ and $(r/r_{200})\leq 0.2$ radial bins. 
On the other hand, the net number of red dwarfs decrease by 20\%. For the 
giants we measure a decrease of 58\% for the blue galaxies and a 7.1\% 
{\it increase} for the red galaxies.

\begin{figure}
\figurenum{5}
\epsscale{1.0}
\plotone{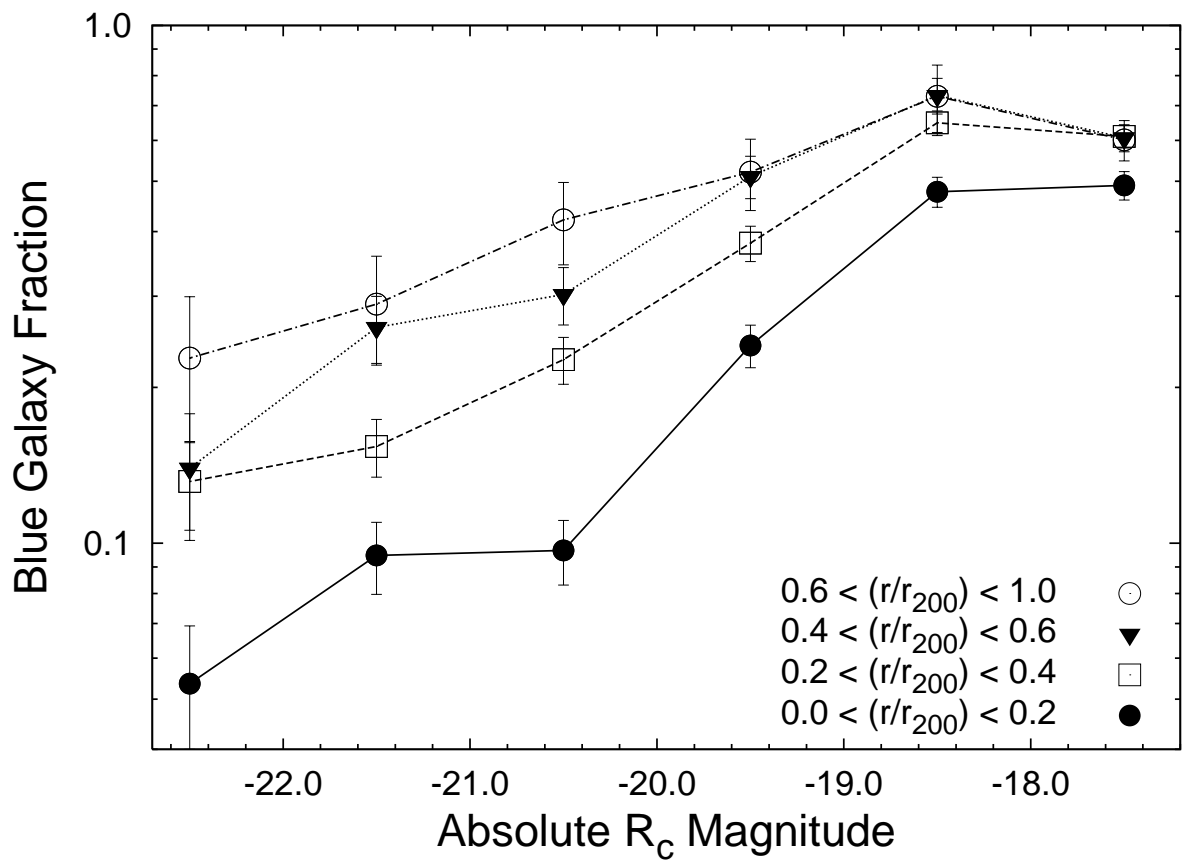}
\caption{$f_{b}$ as a function of magnitude ($M_{R_{c}}$) for 
four radial bins; $(r/r_{200})\leq 0.2$ (filled circles 
with solid line), $0.2<(r/r_{200})<0.4$ (open squares with dashed line), 
$0.4<(r/r_{200})<0.6$ (filled triangles with dotted line), and 
$0.6<(r/r_{200})<1.0$ (open circles with dashed-dotted line).}
\label{Cumlativefb-Radius}
\end{figure}

To ascertain the relative change in the radial-dependence of $f_{b}$ as a 
function of magnitude, we present in Figure~\ref{Cumlativefb-Radius} the 
$f_{b}$ as a function of $M_{R_{c}}$ for the four radial bins used in 
Figure~\ref{fb-Radius} ($f_{b}$ is constructed by counting galaxies in bins 
of one magnitude in width). This figure demonstrates that $f_{b}$ for the 
outer-most annulus ($0.6\leq (r/r_{200})\leq 1.0$) is greater at each magnitude 
interval than for $f_{b}$ measured for the inner annuli. In addition, $f_{b}$ 
for all four radial bins generally show an increase when counting galaxies from 
progressively fainter magnitude bins. The data depicted in 
Figure~\ref{Cumlativefb-Radius} suggests that luminous galaxies may undergo a 
more rapid change in their color composition than the dwarf galaxies. For example, 
$f_{b}$ for galaxies in the $M_{R_{c}}=-22.5$ mag bin declines by a factor of 
4.2 from the outer to the inner radial bin, while $f_{b}$ decreases by a factor 
of 1.5 for galaxies in the $M_{R_{c}}=-18.5$ mag bin ($\sim 2.3\sigma$ difference 
in both cases).

As a caveat we note that our results based on Figures~\ref{fb-Radius} and 
\ref{Cumlativefb-Radius} demonstrate that the counting aperture and magnitude 
range has a direct impact on the value of $f_{b}$, and thus one must be cautious 
when comparing blue fractions for clusters with disparate masses from a variety 
of sources \citep[see also,][]{Ellingson01,Fairley02,DePropris04,Popesso07}.

\subsection{Redshift Dependence of $f_{b}$}

\begin{figure}
\figurenum{6}
\epsscale{1.0}
\plotone{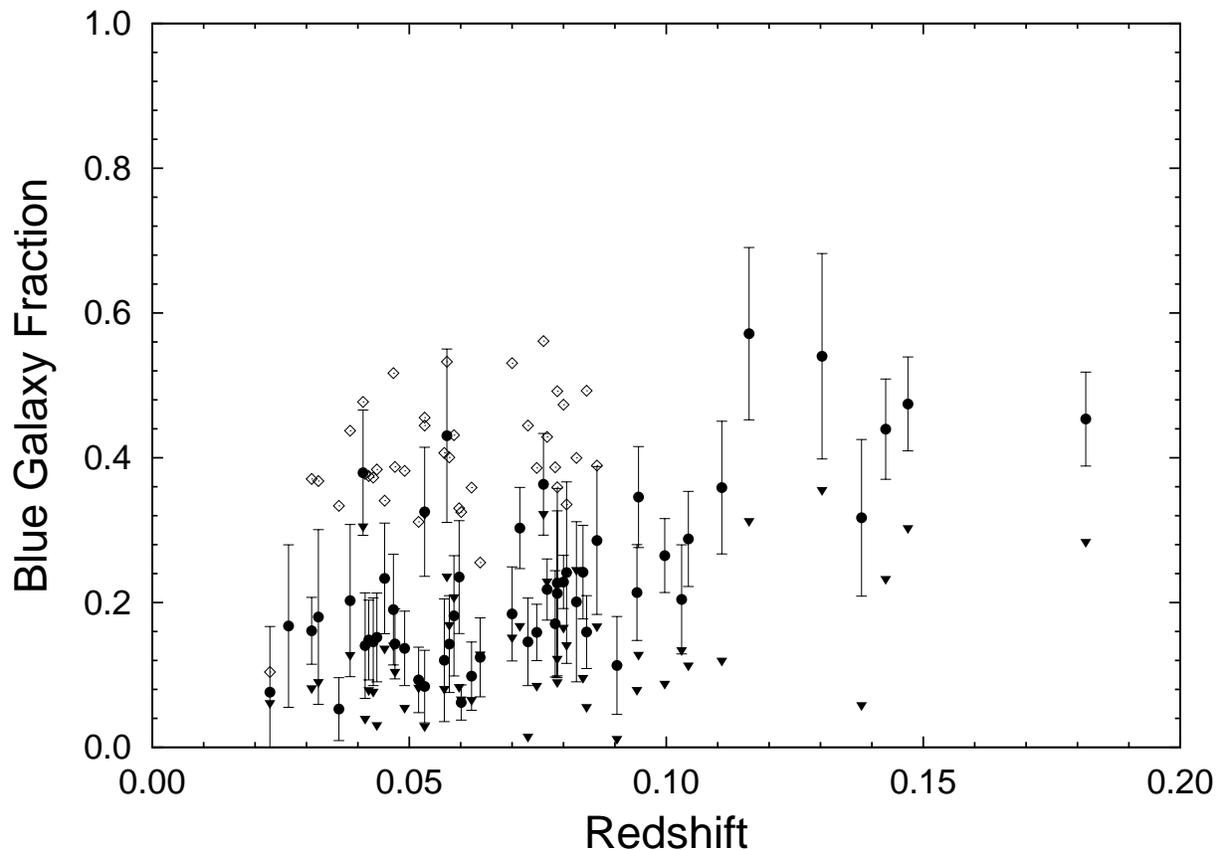}
\caption{Redshift distribution of $f_{b}$ for the cluster galaxy 
population that is 100\% photometrically complete to $M_{R_{c}}=-19$ 
(filled circles), $M_{R_{c}}=-17$ (open diamonds), and $M_{R_{c}}=-20.5$ 
(filled triangles). Error bars for the open symbols are similar to those 
for the filled symbols and have been omitted for clarity. The $f_{b}$ has been 
measured for galaxies within a radius of $(r/r_{200})=0.4$.}
\label{fb-Redshift}
\end{figure}

To test for a correlation between $f_{b}$ and redshift, we present 
in Figure~\ref{fb-Redshift} the $f_{b}$ versus redshift for; a) galaxies 
brighter than $M_{R_{c}}=-19$ (filled symbols; 54 clusters), b) galaxies 
selected with $M_{R_{c}}\leq -17$ (open symbols; 39 clusters), and c) galaxies 
brighter than the equivalent \citet{Butcher84} magnitude limit 
($M_{R_{c}}=-20.5$; solid triangles, 54 clusters). The $f_{b}$'s  
depicted in Figure~\ref{fb-Redshift} are constructed by including only 
galaxies within a clustercentric radius of $(r/r_{200})=0.4$. A Kendell's $\tau$ 
statistic for the $M_{R_{c}}\leq-19$ sample yields a probability of 100\% 
that $f_{b}$ and redshift are correlated. Inspection of 
Figure~\ref{fb-Redshift} shows that $f_{b}$ increases with redshift for 
$z\gtrsim 0.1$. Restricting our analysis to clusters with $z<0.075$, we find a 
correlation probability of only 8\%. Thus most of the correlation between $f_{b}$ and 
redshift is due to clusters with $z\gtrsim 0.1$. For the cluster sample with 
$M_{R_{c}}\leq -17$, we find that $f_{b}$ and redshift are correlated at the 90\% 
significance level. Limiting the $M_{R_{c}}\leq-19$ sample to the same redshift 
range ($z\leq 0.0865$), we find a correlation probability of 98\%. Using the 
magnitude limit $M_{R_{c}}=-20.5$ for the full redshift range ($z<0.2$), we find that 
$f_{b}$ and redshift are correlated at the 97\% significance level.

The correlation between $f_{b}$ and redshift is most-likely a reflection of the 
Butcher-Oemler effect \citep{Butcher78,Butcher84}, in which the fraction of blue 
cluster galaxies increases with look-back time. Since our cluster sample only 
extends to $z\sim0.2$, we are not able to make any firm conclusions on the redshift 
evolution of $f_{b}$. However, Figure~\ref{fb-Redshift} and our correlation 
measurements indicate that the Butcher-Oemler effect is magnitude-dependent, such that 
a fainter magnitude limit yields a larger effect. Unfortunately our $M_{R_{c}}\leq -17$ 
sample lacks a statistically significant number of clusters at $z>0.1$. 

\section{Comparison with Other Results}

\subsection{Dynamical versus Fixed Clustercentric Radius}

A major goal of this paper is to examine the luminosity and color 
distribution of individual and composite cluster galaxy populations. 
Many potential correlations may be obscured when only a limited number 
of cluster galaxies are available. To mitigate this effect, we compiled 
composite samples by stacking together galaxies from individual clusters. 
To minimize radial sampling bias, we scaled each cluster's 
counting aperture by $r_{200}$ prior to combining galaxy counts. 
Radial sampling bias can be problematic when comparing individual clusters (see, 
for example, B07), hence the need to scale clusters by a common dynamical radius. 
This point is aptly illustrated by the recent studies of \citet{Popesso05,Popesso06} 
in which significant correlations between the DGR and various cluster 
characteristics (mass, velocity dispersion, X-ray luminosity, and optical 
luminosity) were found when measuring the DGR using a fixed metric aperture, 
but were much less significant when scaling the counting aperture by $r_{200}$. 
In a study by \citet{Margoniner01} and \citet{Goto03}, cluster richness and $f_{b}$ 
were found to be correlated such that poor systems have a higher $f_{b}$, which 
we suggest is the result of using a fixed counting aperture.

\subsection{Dependence on Magnitude Limits}

Our results, along with those from many others, show that quantities such as the 
DGR and $f_{b}$ are dependent on the luminosity definitions of the galaxy samples 
used. When comparing different studies, care should be taken to account for any 
possible effects arising from the galaxy luminosity or mass limits. Some recent 
studies have examined the DGR and $f_{b}$ based on spectroscopic data from the 
SDSS galaxy sample \citep[e.g.,][]{Aguerri07,Popesso07,Sanchez08}. However, these studies, 
while statistically more robust, are in general much shallower than investigations 
applying a statistical background correction method to photometric data. Studies 
using photometric galaxy samples, going typically two to three magnitudes deeper, 
provide considerably larger leverage in sampling the dependence of galaxy 
population and evolution on luminosity/mass.

\subsection{Dwarf-to-Giant Ratio}

In a study by \citet{DePropris03} the DGR for a spectroscopically-measured 
sample of 60 clusters from the 2dF Galaxy Redshift Survey were presented. 
Transforming their magnitudes to $R_{C}$ \citep{Fukugita95} and adopting our 
cosmology, De Propris et al. defines giants as those galaxies with 
$-24.4\leq M_{R_{c}}\leq -20.9$ and dwarfs with $-20.9\leq M_{R_{c}}\leq -18.9$. 
They divide their galaxies based on spectroscopic type, and find 
$\mbox{DGR}=1.29\pm 0.16$ and $4.88\pm 0.97$ for galaxies with early- and 
late-type spectra, corresponding to our red and blue galaxy samples. Their results 
show the same trend as ours for red and blue galaxies (\S 3.1). The blue DGR 
obtained by De Propris et al. is smaller than the value we measured, and is likely 
due to their division between the giant and dwarf populations being almost 2 mag brighter. 

We also note that De Propris et al. counts galaxies within an aperture of 
$3h^{-1}_{50}$ Mpc in radius rather than scaling relative to a dynamical radius. 
They also determined that the DGR is smaller for galaxies selected within 
$0.6h^{-1}_{50}$ Mpc of the composite cluster center as compared to galaxies at 
larger radii ($2.2\sigma$ difference). In Figure~\ref{RedBlue-Radius} we showed 
that the DGR increases with clustercentric radius for the red+blue cluster 
galaxy population (at the $8\sigma$ level). We suggest that the 
De Propris et al. result would be of higher significance if the cluster 
counting aperture was scaled by a common dynamical radius.

In addition, De Propris et al. reports no statistically significant 
correlation ($\lesssim 1\sigma$) between cluster velocity dispersion (divided 
at $\sigma=800~\mbox{km}~\mbox{s}^{-1}$) and the DGR, or between ``rich'' and 
``poor'' clusters. This result is consistent with our data depicted in 
Figure~\ref{All-DGR-Bgc}, even though we sample 2 mag deeper than De Propris et al. 
and scale by $r_{200}$.

From a study of 69 clusters selected from the RASS-SDSS catalog, 
\citet{Popesso06} presented the DGR for various cluster galaxy sub-samples. 
Using $r_{200}$ as a scaling factor, Popesso et al. found that the DGR is 
not correlated with cluster mass (i.e., $M_{200}$), velocity dispersion, or 
$L_{X}$. This result is compatible with our findings depicted in Figure~\ref{All-DGR-Bgc}, 
where we find no significant correlation between the DGR and 
$B_{gc}$ for the red+blue, red, and blue galaxy populations. Using the $L_{X}$ 
(0.1-2.4 keV) measurements compiled by \citet{Ebeling96,Ebeling00}, we examined 
the DGR vs. $L_{X}$ distribution for galaxies selected within $(r/r_{200})\leq 0.4$. 
Applying the Kendell's $\tau$ statistic to the red+blue/red/blue galaxy samples, we find 
a 87\%/91\%/29\% probability that the DGR and $L_{X}$ are correlated. Our findings support 
the results of Popesso et al. in that there is no strong, statistically significant 
correlation between the DGR and $L_{X}$.

\subsection{Blue Galaxy Fraction}

In addition to examining the DGR for the composite cluster galaxy population, 
Popesso et al. divided their sample into red and blue galaxies by adopting  
$u-r=2.22$ as the dividing color threshold. Popesso et al. found that the 
fraction of red and blue dwarf galaxies decreases toward the cluster 
center (see their Figure 12a). Although they plot the cumulative fractional 
change in the number of dwarfs, it is apparent that Popesso et al. detects 
a more statistically significant drop in the fraction of blue dwarf 
galaxies than what we find (see Figure~\ref{RedBlue-Radius}). This 
difference may be related to their utilization of the $u-$band for the 
selection of blue galaxies and differences in the magnitude range used 
to define giants and dwarfs. 

In Figure 12b from Popesso et al., the dwarf red-to-blue ratio (RBR) is 
depicted as a function of clustercentric radius, normalized to $r_{200}$. 
This figure shows that the RBR is approximately constant from 
$0.4<(r/r_{200})<1.0$ (RBR$\sim0.6$) and increases to $\sim2.4$ near the 
cluster center. Comparison with our Figure~\ref{fb-Radius} shows that 
our $f_{b}$ is consistent with this result. Computing RBR for our dwarf 
sample yields $\mbox{RBR}\sim0.6$ for $(r/r_{200})>0.4$ and $\sim1.4$ 
for the inner-most annulus. These values are similar to those found by 
Popesso et al. given their smaller radial bins.

In a recent study, \citet{Aguerri07} presented the blue galaxy fraction 
for a sample of 88 clusters ($z<0.1$) selected from the SDSS-DR4 
data set. The blue galaxy fraction was constructed by including galaxies 
brighter than $M_{r}=-20$ and located within a radius of 
$(r/r_{200})=1$. Aguerri et al. found that $f_{b}$ is correlated with 
$L_{X}$ in the sense that low $f_{b}$ clusters have a greater $L_{X}$ 
($3\sigma$ difference). This result is also supported by 
\citet{Popesso07}, who measured $f_{b}$ for a sample of 79 clusters 
from the RASS-SDSS cluster catalog by including spectroscopically-detected 
galaxies within $(r/r_{200})=1$. 

Using X-ray data from Ebeling et al., we find that $f_{b}$ and $L_{X}$ 
are not significantly correlated. For galaxies brighter than $M_{R_{c}}=-17$ 
and located within $(r/r_{200})=0.4$, a Kendall's $\tau$ statistic yields 
a 17\% probability of a correlation. Restricting our analysis to $M_{R_{c}}\leq -19$, 
we find a 65\% probability of a correlation. Transforming the magnitude limit utilized 
by Aguerri et al. to our filter and distance scale ($M_{R_{c}}\sim-21$), we find 
a 80\% probability of a correlation. Our results are in agreement with 
\citet{Fairley02} and \citet{Wake05}, who found no significant correlation between 
$f_{b}$ and $L_{X}$.

Utilizing a sample of 60 clusters ($z<0.11$) selected spectroscopically 
from the 2dFGRS, \citet{DePropris04} searched for correlations between 
$f_{b}$ and various cluster properties. In Figure~\ref{Cumlativefb-Radius} 
we showed that $f_{b}$ is sensitive both to the adopted absolute magnitude 
range and clustercentric distance used to select galaxies. De Propris et al. 
reached a similar conclusion by determining that $f_{b}$ increases both 
with decreasing luminosity and increasing clustercentric radius. In addition, 
De Propris et al. also found that there is no significant correlation 
between $f_{b}$ and cluster richness when measured within $(r/r_{200})=0.5$. 
This is also in agreement with our results depicted in Figure~\ref{fb-Bgc}.

\section{Discussion}

In this study we have examined the radial dependence of the luminosity and color 
distribution of cluster galaxies by utilizing the DGR and $f_{b}$. Scaling the 
galaxy counting aperture relative to $r_{200}$, allows us to minimize radial 
sampling bias. 

The main results highlighted in this paper and encapsulated in Figures~\ref{RedBlue-Radius} 
and \ref{fb-Radius}, suggests that some type of dynamical mechanism may be responsible for 
the decline in the number of blue dwarf galaxies relative to the corresponding red systems in 
the cluster core region. The trend depicted in Figure~\ref{RedBlue-Radius} implies that the 
DGR for the blue galaxies is approximately constant with radius. This suggests that the 
decrease in the number of blue dwarfs toward the cluster center is accompanied by a decline in 
the number of blue giants, thus maintaining a roughly constant DGR. A decrease in $f_{b}$ for 
the giant galaxies in the cluster core region (Figure~\ref{fb-Radius}) and a drop in the DGR 
for the red galaxies, suggests that the relative fraction of red giants increases toward the 
cluster center. 

These results support the general view that blue galaxies dominate the galaxy population in 
the outskirts of clusters in contrast to the central cluster region 
\citep[e.g.,][]{Ellingson01,Fairley02,Dahlen04,Tran05}. They also imply that field galaxies, 
which are generally bluer than cluster galaxies \citep[see, for example,][]{Lewis02,McIntosh04}, 
fall into the cluster environment, turn red (possibly via some process that truncates star 
formation), and that blue dwarf galaxies get preferentially disrupted or transformed into red 
dwarfs at small clustercentric radii.  

\subsection{Possible Interpretations}

These findings are open to several possible interpretations: a) blue and red dwarfs get 
disrupted tidally or undergo mergers with giant galaxies at roughly the same rate, which 
destroy individual dwarfs, except in the cluster core region, where the red dwarfs have a 
higher survival rate; b) the average dwarf galaxy star formation rate remains relatively 
unchanged until the dwarfs reach the central cluster region, where the influence of ram 
pressure and cluster tidal effects are expected to be maximized \citep[e.g.,][]{Moore96}, 
quenching star formation and transforming the galaxies into red dwarfs; and c) the 
transformation rate of blue into red galaxies resulting from the quenching of star formation 
is more efficient in giants than in dwarfs (see Figure~\ref{Cumlativefb-Radius}).

The first interpretation requires that blue dwarfs are more susceptible than red dwarfs to 
destructive forces in the cluster central region. Some support for this idea is garnered from the 
fact that blue dwarfs are very similar to the low-mass dwarf spheroidal galaxies, which are 
expected to undergo tidal disruption in the cluster environment 
\citep{Thompson93,Gallagher94,Moore99,Quilis00,Boyce01,Barai07}. These galaxies may potentially 
be the source of tidally-disrupted material that helped to form the halo of cD galaxies 
\citep{Lopez97b,Hilker99,Hilker03}. The red dwarf galaxies, however, may be part of a population 
of nucleated dwarfs \citep{vanDenBergh86,Caldwell87,Lisker07} that would be expected to have a 
deeper gravitational potential well than the more diffuse dwarf spheroidal population. This 
would allow them to more efficiently survive cluster tidal forces against disruption than the 
dwarf spheroidals \citep[i.e., non-nucleated dwarf galaxies; see, for example,][]{Thompson93,Trujillo02,Barai07,
Lisker07}. Nucleated cluster dwarf galaxies have been shown to have colors that are redder on 
average than non-nucleated dwarfs \citep[e.g.][]{Caldwell87,Lisker07}, and thus supports the 
suggestion that the red dwarf population is composed mainly of nucleated dwarf galaxies.

For the second scenario, if the blue dwarf galaxies in the cluster core have been stripped of 
their gas, had their star formation truncated, and transformed into red dwarf galaxies, we 
would expect that the number of red dwarf galaxies would increase with decreasing clustercentric 
radius. Figure~\ref{RedBlue-Radius} indicates that the red DGR actually decreases by a factor of 
$\sim2$ toward the inner-cluster region. Unless a large fraction of blue 
giants, as compared to blue dwarfs, get transformed into red systems, this 
simple explanation for the decreasing number of blue dwarfs in the core does 
not seem plausible. Furthermore, Figure~\ref{RedBlue-Radius} indicates 
that the blue DGR is approximately constant with radius, contrary to what is expected if a large 
fraction of blue giants are transformed into red galaxies in order to decrease the red DGR in the 
cluster core.

The third explanation supposes that the quenching of star formation is more efficient in giant 
galaxies than in dwarfs. Naively, this conjecture can be supported from the results in 
Figure~\ref{Cumlativefb-Radius}, where the more luminous galaxies show a larger decrease in 
$f_{b}$ from the outer- to the inner-cluster region than the less luminous galaxy population. 
However, this seems an unlikely explanation, since, with their lower gravitational potential, 
we would expect most physical mechanisms in clusters that affects star formation rate in 
galaxies should have a relatively larger effect in dwarf galaxies than the more massive giant 
galaxies.

\subsection{Galaxy Disruption versus Fading}

The tidal disruption of dwarf galaxies is a possible physical mechanism to help explain the 
observations presented in this paper; an alternative interpretation is that the blue dwarf galaxies simply 
fade and turn red as they fall toward the cluster center, and are subsequently detected as red galaxies 
in the inner-cluster region. This idea is not unreasonable if we expect that star formation for infalling 
dwarf galaxies gets truncated, with an ensuing passive evolution of the stellar population 
\cite[e.g.,][]{Abraham96,Ellingson01,Treu03,Smith06}. 

The exact number of blue dwarfs that would be expected to fade and turn red, or to disappear due to 
disruption or merger, cannot be accurately calculated 
with a simple toy model. A detailed N-body simulation that incorporates a complete accounting of stellar 
evolution and traces the evolutionary path of each dwarf galaxy would be required. Nevertheless, we are 
able to place limits on the fraction of blue dwarfs that have faded and turned red or that have been 
disrupted, by comparing the blue-to-red dwarf luminosity ratio between the inner- ($L^{i}_{b}/L^{i}_{r}$) 
and outer-most ($L^{o}_{b}/L^{o}_{r}$) radial bins.  

For the case of pure disruption, $L^{o}_{b}/L^{o}_{r}$ was measured for the $0.6\leq (r/r_{200})\leq 1.0$ 
radial bin using background-corrected net galaxy counts. The luminosity of the blue and red galaxies 
were calculated by multiplying the number of galaxies per magnitude bin with the luminosity-equivalent 
of the mid-bin magnitude. Using our magnitude definition of a dwarf galaxy (i.e., $-19.5\leq M_{R_{c}}\leq -17.0$), 
we find that $L^{o}_{b}/L^{o}_{r}=1.81$.

For the inner annulus, $(r/r_{200})\leq 0.2$, $L^{i}_{b}/L^{i}_{r}$ was calculated in the same manner as 
for the outer radial bin except that the magnitude distribution of the background-corrected galaxies is 
determined using the deprojected cluster LF (see B07). The deprojected LF is constructed by subtracting 
the contribution of galaxies located in the cluster outskirts that are projected onto the central cluster 
region. The deprojected LF thus provides a more accurate estimate of the galaxy luminosity distribution 
in the cluster center, especially at the faint end. Using the deprojected LF, we found that 
$L^{i}_{b}/L^{i}_{r}=0.34$. Using our measured luminosity ratios, the expected fraction of disrupted 
blue dwarfs ($f$) can be estimated from $L^{i}_{b}/L^{i}_{r}=(1-f)(L^{o}_{b}/L^{o}_{r})$. Solving for 
$f$ we find $f=0.81$, and thus approximately 81\% of the blue dwarfs would undergo disruption as they 
fall into the central cluster region. 

For the case of pure fading with an associated reddening, we employ the library of evolutionary stellar 
population synthesis models computed using the isochrone code of \citet[][]{Bruzual03}, to provide 
an estimate of the amount of fading/reddening that an infalling galaxy would experience. Adopting the 
concordance cosmological parameters (i.e., $\Omega_{m}=0.3$, $\Omega_{\lambda}=0.7$, and 
$H_{0}=70\,h_{70}~\mbox{km}~\mbox{s}^{-1}~\mbox{Mpc}^{-1}$), the timescale for the infall of 
a typical dwarf galaxy from the $0.6\leq (r/r_{200})\leq 1.0$ annulus is estimated by adopting 
the average value of $r_{200}$ and the mean velocity dispersion for our 57-cluster sample. 
Using the $B_{gc}$ values tabulated in B07, and the relationship between $B_{gc}$ and velocity 
dispersion from \citet{Yee03}, we find $\overline{r_{200}}=1.6\,h^{-1}_{70}$ Mpc and 
$\overline{\sigma_{v}}=840\,\mbox{km}~\mbox{s}^{-1}$. Using these values yields an average infall 
timescale of approximately 2 Gyr. Adopting a Salpeter Initial Mass Function \citep{Salpeter55}, 
solar metallicity, a single-burst star formation model to simulate star formation truncation, 
and a 2 Gyr time-frame for passive evolution, we predict a $\sim0.2$ mag fading in $R_{C}$ and 
a reddening of $\Delta(B-R_{c})\sim0.5$ mag.

For the case of fading/reddening of dwarf galaxies, we invoke a similar procedure as that used for 
the disruption scenario. For this case we count dwarf galaxies in the $0.6\leq (r/r_{200})\leq 1.0$ 
annulus using the magnitude range $-19.7\leq M_{R_{C}}\leq -17.2$ in order to compensate, to 
first order, for the expected 0.2 mag fading for an infalling galaxy. For the outer radial bin we 
find $L^{o}_{b}/L^{o}_{r}=1.46$. The expected fraction of galaxies that have undergone fading with 
an associated reddening can be estimated from 
$L^{i}_{b}/L^{i}_{r}=(1-f)L^{o}_{b}/(L^{o}_{r}+fL^{o}_{b})$. Solving for $f$ yields $f=0.57$, and 
thus approximately 57\% of the blue dwarfs would be expected to have undergone fading and reddening 
as they reach the inner-cluster region. 

It seems unlikely, however, that a pure fading scenario can explain the change of $f_b$ seen from 
the outer to inner region of clusters. First, applying the same procedure of fading to the giant 
galaxies, we find that 80\% of the blue galaxies are expected to undergo fading and reddening by 
the time they reach the cluster core from the outer annulus. This is likely the dominant cause for 
the change in $f_b$ for the giant galaxies, since it is unlikely that they can be easily destroyed 
by tidal forces. If we assume that blue dwarfs were to fade by the same fraction, then the dwarf 
blue-to-red luminosity would be much smaller than what is measured, by a factor of about $\sim2.5$; 
unless some mechanism is invoked that fades blue giant galaxies by a much larger fraction than 
dwarf galaxies. Due to the lower gravitational potential possessed by dwarf galaxies, almost all 
mechanisms used to explain the hastening of galaxy evolution in rich environments operate equally 
or more efficiently for lower-mass galaxies; e.g., ram pressure, tidal interactions, harassment, etc. 
The only mechanism that could produce a higher fraction of the quenching of star formation in 
massive galaxies is AGN feedback, in that more massive galaxies will be more likely to contain a 
massive black hole required for the AGN activity.

Second, in a pure fading scenario for dwarf galaxies, we would expect the DGR for the blue+red sample 
to stay approximately constant from the outer to the inner region, unless the parent populations of 
galaxies from which the inner and outer regions are drawn from are very different. Instead, we find 
there is a factor of 4 difference in the DGR between the outer annulus and the cluster core.

While the discussion above cannot completely rule out the pure fading scenario, it seems likely that, 
as blue dwarf galaxies fall into the cluster core, at least some fraction of them will be destroyed 
either by tidal disruption, or mergers with larger galaxies.

\section{Conclusions}

In this paper we have studied the luminosity and color properties of a sample 
of 57 low-redshift Abell clusters. Our main conclusions are:

1) The DGR for the red, blue, and red+blue cluster galaxies are independent of cluster 
richness when scaling the counting aperture by a dynamical radius 
(i.e., $r_{200}$). Also, the DGR for blue galaxies is larger than for red 
systems.

2) The DGR for the red galaxies decreases in the inner cluster region, while 
the blue DGR is approximately constant as a function of cluster-centric radius.

3) The $f_{b}$ was found not to correlate with cluster richness 
when counting galaxies within a dynamical radius; however, it is found 
to be correlated with the adopted counting aperture and magnitude limit.

4) The $f_{b}$ for dwarf galaxies was found to be approximately constant 
with clustercentric radius except in the cluster core region where $f_{b}$ 
decreases. 

5) The $f_{b}$ for giant galaxies was found to increase with clustercentric radius 
for all measured annuli.

6) Based on the clustercentric radial dependence of the DGR and the galaxy blue fraction, 
it is unlikely that either a pure disruption or a pure fading/reddening scenario can 
describe the evolution of infalling dwarf galaxies; both outcomes are produced by the 
cluster environment.

\acknowledgments

We thank the anonymous referee for reviewing our paper. 
Research by W. A. B. is supported by a start-up grant from the University of 
North Dakota. Research by H. K. C. Y. is supported by an NSERC Discovery grant. 
O. L.-C research is supported by INAOE and a CONACyT grant for Ciencia B{\'a}sica 
P45952-F. O. L.-C. acknowledges support from a research grant from the 
Academia Mexicana de Ciencias-Royal Society during 2006-2007 taken to the 
University of Bristol. OLC acknowledges Gus Oemler for suggesting the use of 
non-parametric DGR. We thank Huan Lin for providing photometric catalogs 
for five control fields, and James Brown for the use of his galaxy profile 
fitting software and photometric data for A496 and A1142.

The Image Reduction and Analysis Facility (IRAF) is distributed by the 
National Optical Astronomy Observatory, which is operated by AURA, Inc., 
under contract to the National Science Foundation. This research has made 
use of the NASA/IPAC Extragalactic Database (NED) which is operated by the 
Jet Propulsion Laboratory, California Institute of Technology, under 
contract with the National Aeronautics and Space Administration.


\clearpage

\clearpage

\clearpage

\clearpage

\clearpage

\clearpage

\clearpage

\begin{thebibliography}{}

\bibitem[Abadi et~al.(1999)]{Abadi99} Abadi, M. G., Moore, B., \& 
Bower, R. G. 1999, \mnras, 308, 947

\bibitem[Abraham et~al.(1996)]{Abraham96} Abraham, R. G., et al. 1996, 
\apj, 461, 694

\bibitem[Aguerri et~al.(2007)]{Aguerri07} Aguerri, J. A. L., 
S\'anchez-Janssen, R., \& Mu\~{n}oz-Tu\~{n}\'on, C. 2007, \aap, 471, 17

\bibitem[Barai et~al.(2007)]{Barai07} Barai, P., Brito, W., \& Martel, H. 
2007, preprint (astro-ph/0707.1533)

\bibitem[Barkhouse(2003)]{Barkhouse03} Barkhouse, W. A. 2003, Ph.D. Thesis, 
Univ. Toronto

\bibitem[Barkhouse et~al.(2007)]{Barkhouse07} Barkhouse, W. A., Yee, 
H. K. C., \&  L\'opez-Cruz, O. 2007, \apj, 671, 1471

\bibitem[Boselli \& Gavazzi(2006)]{Boselli06} Boselli, A., \& Gavazzi, G. 
2006, \pasp, 118, 517

\bibitem[Boyce et~al.(2001)]{Boyce01} Boyce, P. J., Phillipps, S., 
Jones, J. B., Driver, S. P., Smith, R. M., \& Couch, W. J. 2001, \mnras, 
328, 277

\bibitem[Brown(1997)]{Brown97} Brown, J. P. 1997, Ph.D. thesis, Univ. Toronto

\bibitem[Bruzual \& Charlot(2003)]{Bruzual03} Bruzual, G., \& Charlot, S. 
2003, \mnras, 344, 1000

\bibitem[Burstein \& Heiles(1982)]{Burstein82} Burstein, D., \&  Heiles, C. 
1982, \aj, 87, 1165

\bibitem[Burstein \& Heiles(1984)]{Burstein84} Burstein, D., \& Heiles, C. 
1984, \apjs, 54, 33

\bibitem[Butcher \& Oemler(1978)]{Butcher78} Butcher, H., \& Oemler, A. 
1978, \apj, 219, 18

\bibitem[Butcher \& Oemler(1984)]{Butcher84} Butcher, H., \& Oemler, A. 
1984, \apj, 285, 426

\bibitem[Caldwell \& Bothun(1987)]{Caldwell87} Caldwell, N., \& Bothun, 
G. D. 1987, \aj, 94, 1126

\bibitem[Christlein \& Zabludoff(2003)]{Christlein03} Christlein, D., \& 
Zabludoff, A. I. 2003, \apj, 591, 764

\bibitem[Cole \& Lacey(1996)]{Cole96} Cole, S., \& Lacey, C. 1996, \mnras, 
281, 716

\bibitem[Dahl\'en et~al.(2004)]{Dahlen04} Dahl\'en, T., Fransson, C., 
\"{O}stlin, G., \&  N\"{a}slund, M. 2004, \mnras, 350, 253

\bibitem[De~Propris et~al.(2003)]{DePropris03} De Propris, R., et al. 
2003, \mnras, 342, 725

\bibitem[De~Propris et~al.(2004)]{DePropris04} De Propris, R., et al. 
2004, \mnras, 351, 125

\bibitem[Dressler(1980)]{Dressler80} Dressler, A. 1980, \apj, 236, 351

\bibitem[Dressler et~al.(1997)]{Dressler97} Dressler, A., et al. 1997, 
\apj, 490, 577

\bibitem[Driver et~al.(1998)]{Driver98} Driver, S. P., Couch, W. J., \& 
Phillipps, S. 1998, \mnras, 301, 369

\bibitem[Dubinski(1998)]{Dubinski98} Dubinski, J. 1998, \apj, 502, 141

\bibitem[Ebeling et~al.(1996)]{Ebeling96} Ebeling, H., Voges, W., 
B\"{o}hringer, H., Edges, A. C., Huchra, J. P., \& Briel, U. G. 1996, 
\mnras, 281, 799

\bibitem[Ebeling et~al.(2000)]{Ebeling00} Ebeling, H., Edges, A. C., 
Allen, S. W., Crawford, C. S., Fabian, A. C., \& Huchra, J. P. 2000, 
\mnras, 318, 333

\bibitem[Ellingson et~al.(2001)]{Ellingson01} Ellingson, E., Lin, H., 
Yee, H. K. C., \& Carlberg, R. G. 2001, \apj, 547, 609

\bibitem[Fairley et~al.(2002)]{Fairley02} Fairley, B. W., Jones, L. R., 
Wake, D. A., Collins, C. A., Burke, D. J., Nichol, R. C., \&  Romer, A. K. 
2002, \mnras, 330, 755

\bibitem[Fukugita et~al.(1995)]{Fukugita95} Fukugita, M., Shimasaku, K., 
\& Ichikawa, T. 1995, \pasp, 107, 945

\bibitem[Gallagher \& Wyse(1994)]{Gallagher94} Gallagher, J. S., \& Wyse, 
R. F. G. 1994, \pasp, 106, 1225

\bibitem[Goto et~al.(2003)]{Goto03} Goto, T., et al. 2003, \pasj, 55, 739

\bibitem[Hansen et~al.(2005)]{Hansen05} Hansen, S. M., McKay, T. A., 
Wechsler, R. H., Annis, J., Sheldon, E. S., \& Kimball, A. 2005, 
\apj, 633, 122

\bibitem[Hilker et~al.(1999)]{Hilker99} Hilker, M., Infante, L., \& 
Richtler, T. 1999, \aaps, 138, 55

\bibitem[Hilker et~al.(2003)]{Hilker03} Hilker, M., Mieske, S., \& 
Infante, L. 2003, \aap, 397, L9

\bibitem[Jones \& Forman(1999)]{Jones99} Jones, C., \& Forman, W. 1999, 
\apj, 511, 65

\bibitem[Kent \& Gunn(1982)]{Kent82} Kent, S. M., \& Gunn, J. E. 1982, 
\aj, 87, 945

\bibitem[Landolt(1992)]{La92} Landolt, A. U. 1992, \aj, 104, 372

\bibitem[Lewis et~al.(2002)]{Lewis02} Lewis, I., et al. 2002, \mnras, 334, 673

\bibitem[Lisker et~al.(2007)]{Lisker07} Lisker, T., Grebel, E. K., 
Binggeli, B., \& Glatt, K. 2007, \apj, 660, 1186

\bibitem[L\'opez-Cruz(1997)]{Lopez97} L\'opez-Cruz O. 1997, Ph.D. Thesis, 
Univ. Toronto

\bibitem[L\'opez-Cruz et~al.(1997b)]{Lopez97b} L\'opez-Cruz, O., Yee, 
H. K. C., Brown, J. P., Jones. C., \& Forman, W. 1997, \apj, 475, L97

\bibitem[L\'opez-Cruz(2001)]{Lopez01} L\'opez-Cruz, O. 2001, Rev. Mex. 
Astron. Astrophys, Conf. Ser., 11, 183

\bibitem[L\'opez-Cruz et~al.(2004)]{Lopez04} L\'opez-Cruz, O., Barkhouse, 
W. A., \& Yee, H. C. K. 2004, \apj, 614, 679

\bibitem[Margoniner et~al.(2001)]{Margoniner01} Margoniner, V. E., 
De Carvalho, R. R., Gal, R. R., \& Djorgovski, S. G. 2001, \apj, 548, L143

\bibitem[McIntosh et~al.(2004)]{McIntosh04} McIntosh, D. H., Rix, H.-W. \& 
Caldwell, N. 2004, \apj, 610, 161

\bibitem[Merritt(1984)]{Merritt84} Merritt, D. 1984, \apj, 276, 26

\bibitem[Moore et~al.(1996)]{Moore96} Moore, B., Katz, N., Lake, G., 
Dressler, A., \& Oemler, A. 1996, \nat, 379, 613

\bibitem[Moore et~al.(1999)]{Moore99} Moore, B., Lake, G., Quinn, T., 
Stadel, J. 1999, \mnras, 304, 465

\bibitem[Morris et~al.(1998)]{Morris98} Morris, S. L., Hutchings, J. B., 
Carlberg, R. G., Yee, H. K. C., Ellingson, E., Balogh, M. L., Abraham, 
R. G., \& Smecker-Hane, T. A. 1998, \apj, 507, 84
 
\bibitem[Okamoto \& Nagashima(2001)]{Okamoto01} Okamoto, T., \& Nagashima, 
M. 2001, \apj, 547, 109

\bibitem[Popesso et~al.(2005)]{Popesso05} Popesso, P., B\"{o}hringer, H., 
Romaniello, M., \& Voges, W. 2005, \aap, 433, 415

\bibitem[Popesso et~al.(2006)]{Popesso06} Popesso, P., Biviano, A., 
B\"{o}hringer, H., \& Romaniello, M. 2006, \aap, 445, 29

\bibitem[Popesso et~al.(2007)]{Popesso07} Popesso, P., Biviano, A., 
Romaniello, M., \& B\"{o}hringer, H. 2007, \aap, 461, 411

\bibitem[Press et~al.(1992)]{Press92} Press, W. H., Teukolsky, S. A., 
Vetterling, W. T., \& Flannery, B. P. 1992, Numerical Recipes, The 
Art of Scientific Computing, (2d ed.; Cambridge: Cambridge University Press)

\bibitem[Quilis et~al.(2000)]{Quilis00} Quilis, V., Moore, B., \& Bower, R. 
2000, Science, 288, 1617

\bibitem[Rood et~al.(1972)]{Rood72} Rood, H. J., Page, T. L., Kintner, E. C., 
\& King, I. R. 1972, \apj, 175, 627

\bibitem[Roos \& Norman(1979)]{Roos79} Roos, N., \& Norman, C. A. 1979, 
\aap, 76, 75

\bibitem[Salpeter(1955)]{Salpeter55} Salpter, E. E. 1955, \apj, 121, 161

\bibitem[S\'{a}nchez-Janssen et al.(2008)]{Sanchez08} S\'{a}nchez-Janssen, R., 
Aguerri, J. A. L., \& Mu\~{n}oz-Tu\~{n}\'{o}n, C. 2008, \apj, 679, L77

\bibitem[Schechter(1976)]{Schechter76} Schechter, P. 1976, \apj, 203, 297

\bibitem[Smith et~al.(2006)]{Smith06} Smith, R. J., Hudson, M. J., Lucey, 
J. R., Nelan, J. E., \& Wegner, G. A. 2006, \mnras, 369, 1419

\bibitem[Thomas \& Katgert(2006)]{Thomas06} Thomas, T., \& Katgert, P. 
2006, \aap, 446, 31

\bibitem[Thompson \& Gregory(1993)]{Thompson93} Thompson, L. A., \& 
Gregory, S. A. 1993, \aj, 106, 2197

\bibitem[Thuan \& Gunn(1976)]{Thuan76} Thuan, T. X., \& Gunn, J. E. 1976, 
\pasp, 88, 543 

\bibitem[Tran et~al.(2005)]{Tran05} Tran, K.-V. H., van Dokkum, P., 
Illingworth, G. D., Kelson, D., Gonzalez, A., \& Franx, M. 2005, \apj, 
619, 134

\bibitem[Treu et~al.(2003)]{Treu03} Treu, T., Ellis, R. S., Kneib, J.-P., 
Dressler, A., Smail, I., Czoske, O., Oemler, A., \& Natarajan, P. 2003, 
\apj, 591, 53 

\bibitem[Trujillo et~al.(2002)]{Trujillo02} Trujillo, I., Aguerri, J. A. L., 
Guti\'errez, C. M., Caon, N., \& Cepa, J. 2002, \apj, 573, 9

\bibitem[van~den~Bergh(1986)]{vanDenBergh86} van den Bergh, S. 1986, \aj, 
91, 271

\bibitem[Wake et~al.(2005)]{Wake05} Wake, D. A., Collins, C. A., 
Nichol, R. C., Jones, L. R., \& Burke, D. J. 2005, \apj, 627, 186

\bibitem[Wolf et~al.(2007)]{Wolf07} Wolf, C., Gray, M. E., Arag\'on-Salamanca, 
A., Lane, K. P., \& Meisenheimer, K. 2007, \mnras, 376, L1

\bibitem[Yee(1991)]{Yee91} Yee, H. K. C. 1991, \pasp, 103, 396

\bibitem[Yee \& L\'opez-Cruz(1999)]{Yee99} Yee, H. K. C., \& L\'opez-Cruz, 
O. 1999, \aj, 117, 1985

\bibitem[Yee \& Ellingson(2003)]{Yee03} Yee, H. K. C. \&  Ellingson, E. 
2003, \apj, 585, 215


\end{thebibliography}
\end{document}